\documentclass[prl,aps,showpacs,twocolumn, a4paper,tightenlines,balancelastpage]{revtex4}
\usepackage{amsmath,amsfonts,amssymb,mathrsfs,bm}
\usepackage{verbatim,psfrag,times,CJK}
\usepackage{graphicx,graphics,color,epsfig}
\usepackage{flafter}

\newcommand{\ket}[1]{| #1 \rangle}

\begin{document}

\title{Nullity of quantum discord of two-qubit $\emph{X}$-state systems}

\author{Gao-xiang \surname{Li}$^{a}$}
%\email{gaox@phy.ccnu.edu.cn}
\author{Zhen \surname{Yi}$^{a}$}
\author{Zbigniew \surname{Ficek}$^{b}$}

\affiliation{$^{a}$Department of Physics, Huazhong Normal University, Wuhan 430079,
PR China\\
$^{b}$The National Centre for Mathematics and Physics, KACST, P.O. Box 6086, Riyadh 11442, Saudi Arabia}

\begin{abstract}
The conditions are established for nullity of quantum discord for mixed states of a general two-qubit system whose dynamical behaviour is given by an $X$-state form density matrix. It is found that quantum discord can vanish at a finite time even in the presence of correlations between qubits. The degeneracy of the populations of the combined states of the qubits and the multi-qubit quantum coherences are shown to be responsible for vanishing of the discord. We illustrate our considerations by examining the coherent-state Tavis-Cummings model in which two qubits interact dispersively with the cavity field. The disappearance of the discord is shown to be dependent on the initial atomic conditions that it may occur at a discrete instance or periodically at discrete instances redistributed over whole range of the evolution time. The nullity of the discord is interpreted as resulting from equal preparation of the spin anti-correlated and spin correlated product states in superposition~states. 
\end{abstract}

\date{\today}

\maketitle

Nothing in the principles of quantum mechanics prevents a system from possessing quantum correlations. Quantum or nonclassical correlations are unavoidable in quantum systems~\cite{nc00}, and as it has been established recently, almost all quantum states have nonclassical correlations or, equivalently, have positive discord~\cite{oz01,fa10,mp10,ar10,gg10}. Quantum discord, one of the measures of quantum correlations, is closely related to entanglement, a quantum feature of composite systems that cannot be represented as a product of the states of the individual systems, but recent theoretical investigations indicated that this may not be true. Only for pure states, quantum discord is equivalent to entanglement and significant departures between these quantities have been reported for mixed states. For a large class of states, entanglement is zero, whereas the quantum discord is positive for almost all quantum states. Over the last two years, a series of investigations, both theoretical and experimental, has shown that the quantum discord of open quantum systems vanishes asymptotically in time~\cite{xx10}, contrary to entanglement that can disappear at finite time~\cite{ye09}. Understanding and manipulating dynamics of these quantum features are of great importance for both fundamental physics and new emerging quantum technologies.

In principle, it is possible to construct multi-qubit states with zero discord. An explicit construction of such states has recently been investigated in many publications~\cite{dv10,da10,aj10,al10,zz11}. In these investigations conditions for nullity of quantum discord have been established. However, the zero-discord states such constructed suffer from a common handicap: they posses no any correlations and coherences. Therefore, it is not surprising that the discord is zero.

Somewhat less familiar is a construction of states that could have zero discord at a finite time even in the presence of correlations between the qubits. This is the purpose of the present Letter to show that it is really possible to prepare a two-qubit system in a quantum state that can exhibit a vanishing discord at a finite time with non-zero quantum coherences present in the system. In a simple example of the coherent-state Tavis-Cummings model, we show how it is possible to generate such states in a dynamical process and what their properties are. Zero-discord states are particularly attractive regarding the recent prove that vanishing discord is necessary and sufficient for completely positive maps~\cite{sl09}, and no constrains imposed on local broadcasting of correlated states~\cite{ph08}.

We choose a two qubit system whose dynamical behaviour is given by an $X$-state form density matrix, and examine under what circumstances the system can have zero discord with nonzero correlations between the qubits. We illustrate our examinations on a simple model of two two-level atoms interacting dispersively with a damped single-mode cavity field.

Consider a pair of two-level atoms (qubits), labelled by the suffices $A$ and $B$, prepared in a quantum state determined by the $X$-state form density matrix $\rho$. The matrix, written in the space spanned by spin correlated $(\ket 1\equiv \ket{g_{A},g_{B}}, \ket 4\equiv \ket{e_{A},e_{B}})$ and spin anti-correlated $(\ket 2\equiv \ket{g_A,e_{B}}, \ket 3\equiv \ket{e_{A},g_{B}})$ product states is of the form
\begin{eqnarray}
\rho_{AB} = \sum\limits_{j=1}^4\rho_{jj}|j\rangle\langle j| +\left(\rho_{14}|1\rangle\langle 4|+\rho_{23}|2\rangle\langle 3| + {\rm H.c.}\right) ,\label{e1}
\end{eqnarray}
where $\ket{g_{j}}$ and $\ket{e_{j}}\, (j=A,B)$ represent the ground and excited states of the qubits. The state (\ref{e1}) is of the form associated with quantum coherences existing between states of both qubits, not between the states of individual qubits. These coherences tell us about the correlations between the qubits. 

We now determine for which states of a bipartite system, specified by the density matrix of an $X$-state form, the quantum discord can reach zero even in the presence of internal coherences. We define the quantum discord by relation to the quantum mutual information and adopt a procedure which is both analytic and simple, and points out the unusual behaviour of the discord in a straightforward and transparent fashion. 

The total correlations (quantum and classical) in a bipartite quantum system are measured by the total quantum mutual information $I(\rho_{AB})$ defined as
\begin{eqnarray}
I(\rho_{AB}) &=& S(\rho_A) + S(\rho_B) - S(\rho_{AB}) \nonumber\\
&=& D(\rho_{AB})+C(\rho_{AB}) ,\label{e2}
\end{eqnarray}
where $\rho_{A(B)}$ and $\rho_{AB}$ are the reduced density matrices of the subsystem $A(B)$ and the total system, respectively; $S(\rho)=-{\rm Tr}\{\rho\log_2\rho\}$ is the von Neumann entropy, $D(\rho_{AB})$ is the quantum discord which provides information about the quantum nature of the correlations, and $C(\rho_{AB})$ describes the classical correlations, which can be obtained by use of a measurement-based conditional density operator~\cite{oz01,fa10,mp10,ar10}. Naturally, if the mutual information is larger than the classical correlations, the quantum discord is positive, different from zero. Clearly, the quantum discord is obtained by subtracting the classical correlations from the total mutual information.  

The mechanism for a positive discord is a quantum correlation between the $A$ and $B$ subsystems. The subsystems are correlated if the measurement of an observable of the~$A$ system projects the $B$ system into a new state, and vice versa. Here, we limit ourselves to projective measurements performed locally only on the subsystem $B$ described by a complete set of orthonormal projectors $\{\Pi_k\}$ corresponding to the outcomes $k$. The classical correlations $C(\rho_{AB})$ are then defined as
\begin{eqnarray}
C(\rho_{AB}) = \max\limits_{\{\Pi_k\}}[S(\rho_A)-S(\rho_{AB}|\{\Pi_k\}] ,\label{e3}
\end{eqnarray}
where the maximum is taken over the set of projective measurement $\{\Pi_k\}$, and $S(\rho_{AB}|\{\Pi_k\})=\sum_kp_kS(\rho_k)$ is the conditional entropy of the subsystem $A$, with $\rho_k= {\rm Tr}_B(\Pi_k\rho_{AB}\Pi_k)/p_k$ and $p_k= {\rm Tr}_{AB}(\rho_{AB}\Pi_k)$. 

We now link the classical correlations as defined in Eq.~(\ref{e3}) to the quantum discord. We see that a state that could minimize the conditional entropy $S(\rho_{AB}|\{\Pi_k\})$ would then correspond to a zero-discord state~\cite{oz01}. Following this, we introduce the basis of orthogonal states of the subsystem $B$; $|+\rangle=\cos\theta |e_B\rangle+\sin\theta e^{i\phi}|g_B\rangle$ and $|-\rangle=\sin\theta |e_B\rangle-\cos\theta e^{i\phi}|g_B\rangle$. Then, we define a general one-qubit projector $\Pi_k=|k\rangle\langle k|$ $(k=\pm)$ on the subsystem $B$. From this, we find that the minimum value of the conditional entropy  $S(\rho_{AB}|\{\Pi_k\})$, that maximizes the classical correlations, can be analytically expressed~as
\begin{eqnarray}
\min\limits_{\{\Pi_k\}}\{S(\rho_{AB}|\{\Pi_k\})\} = \min\{C_{m1},C_{m2}\} ,
\end{eqnarray}
where $C_{m1}$ results from the projective measurements on $B$, with the angle $\theta=0$ or $\pi/2$:
\begin{align}
&C_{m1} = S(\chi_{m1})-S({\rm Tr}_A(\chi_{m1})) \nonumber\\
&= -\rho_{44}\log_{2}\left(\frac{\rho_{44}}{\rho_{22}+\rho_{44}}\right) -\rho_{22}\log_{2}\left(\frac{\rho_{22}}{\rho_{22}+\rho_{44}}\right) \nonumber\\ 
& -\rho_{33}\log_{2}\left(\frac{\rho_{33}}{\rho_{11}+\rho_{33}}\right) -\rho_{11}\log_{2}\left(\frac{\rho_{11}}{\rho_{11}+\rho_{33}}\right) ,
\end{align}
with the zero-discord state $\chi_{m1}=\sum_{j=1}^4\rho_{jj}|j\rangle\langle j|$.

Accordingly, $C_{m2}$ results from the projective measurements on $B$, with $\theta=\pi/4$ and $\phi = (\phi_{1}-\phi_{2})/2$:
\begin{align}
&C_{m2} = S(\chi_{m2})-S({\rm Tr}_A(\chi_{m2})) \nonumber\\
&=-\frac{(1+\Upsilon)}{2}\!\log_{2}\!\left(\frac{1+\Upsilon}{2}\!\right)
-\frac{(1-\Upsilon)}{2}\!\log_{2}\!\left(\frac{1-\Upsilon}{2}\!\right) ,\label{e6a}
\end{align}
where $\Upsilon = \left[(\rho_{11} + \rho_{22} - \rho_{33} - \rho_{44})^{2} + 4(|\rho_{41}| + |\rho_{32}|)^2\right]^{\frac{1}{2}}$, $\phi_{1}=\arg(\rho_{14})$, $\phi_{2}=\arg(\rho_{23})$ and the zero-discord state 
\begin{align}
\chi_{m2} &= \frac{1}{2}\left\{(\rho_{11}+\rho_{22})(|1\rangle\langle 1|+|2\rangle\langle 2|)\right.\nonumber\\
&+\left. \!(\rho_{33}+\rho_{44})(|3\rangle\langle 3|+|4\rangle\langle 4|)\right. \nonumber\\
&+\left. \!\left(|\rho_{14}|\!+\!|\rho_{23}|\right)\!\!\left[\left(|1\rangle \langle 4|{\rm e}^{i\phi_{1}}\!+\!|2\rangle\langle 3|{\rm e}^{i\phi_{2}}\right)\!+\!{\rm H.c.}\right]\!\right\} .
\end{align}

Evidently, the states $\chi_{m1}$ and $\chi_{m2}$ which posses zero quantum discord have the same $X$-state form as the state $\rho_{AB}$. From this observation, it follows that for a two qubit system determined by the density matrix of an $X$-state form, the quantum discord vanishes when
\begin{eqnarray}
\rho_{14} = \rho_{23} = 0 ,\label{e6}
\end{eqnarray}
or when
\begin{eqnarray}
 \rho_{11} = \rho_{22} ,\quad \rho_{33} = \rho_{44} ,\quad {\rm and}\quad |\rho_{23}| = |\rho_{14}| .\label{e7}
\end{eqnarray}
Thus, we have extracted two distinct conditions for nullity of the quantum discord. 
The former, Eq.~(\ref{e6}), is the condition involving only the diagonal density matrix elements (populations) with all quantum coherences equal to zero. This is the trivial case which suggests that the zero discord occurs when there are no any quantum coherences in the system. From this it is clear that the quantum discord vanishes. It is also clear that outside this condition, the discord is always different from zero. This has already been reported in several publications~\cite{dv10,da10,aj10,al10,zz11}.

The latter condition, Eq.~(\ref{e7}), is more interesting and in fact surprising. It is a general result valid for any initial state and any dynamical process that involves the $X$-state form density matrix. It shows that a zero-discord state does not guarantee {\it no quantum coherences} in the system. The condition to be satisfied requires the pair degeneracy of the populations of the energy states and equal magnitudes of the one- and two-photon quantum coherences. It is easily verified that the condition is related physically to the preparation of the spin correlated $(\ket 1,\ket 4)$ and anti-correlated $(\ket 2,\ket 3)$ states in equal superposition states. Note that in general, the superposition states so obtained are non-maximally entangled states. In other words, they do not coincide with the Bell states.

We now demonstrate the occurrence of the nullity of quantum discord in a practical physical system: The coherent-state Tavis-Cummings model of two atoms coupled to a single cavity mode~\cite{zg00}. The interatomic separation is assumed to be much larger than the transition wavelength so that cooperative effects can be ignored. Furthermore, we consider the atoms to be well localized and neglect their external motion. If we specialise to the case of a highly detuned cavity mode, $|\delta|\gg g$, where $g$ is the coupling constant of the atoms to the cavity field, and $\delta=\omega_0-\omega$ is the detuning between the atomic transition frequency $\omega_0$ and cavity frequency~$\omega$, the effective Hamiltonian of the system reads
\begin{eqnarray}
\emph{H}_{{\rm eff}} &=& \frac{1}{2}\lambda \left[\sum_{j=A,B} (|e_{j}\rangle\langle{e}_{j}|aa^{\dag}-|g_{j}\rangle\langle{g}_{j}|a^{\dag}a)\right. \nonumber\\
&&\left. +\left(\sigma_{-}^{A}\sigma_{+}^{B}+\sigma_{+}^{A}\sigma_{-}^{B}\right)\right] ,\label{e11}
\end{eqnarray}
where $\sigma_+^{j}=|e_j\rangle\langle g_j|$ and $\sigma_-^{j}=|g_j\rangle\langle e_j|\, (j=A,B)$ are atomic spin operators, $a^\dag$ and $a$ are respectively the creation and annihilation operators of the cavity field, $\lambda =g^2/2\delta$ is the effective coupling constant (Rabi frequency). The first term in the Hamiltonian (\ref{e11}) represents the intensity dependent Stark shift, while the second term is of the form analogous to the familiar dipole-dipole interaction between the atoms~\cite{ft02}.

If we include the dissipation of the cavity mode, the dynamics of the system are then determined by the the density matrix $\rho(t)$, which satisfies the following master equation
\begin{eqnarray}
\dot{\rho}(t) = -i[H_{{\rm eff}},\rho(t)]+\kappa{\cal L}_c\rho(t) ,\label{e12}
\end{eqnarray}
where ${\cal L}_c\rho(t)= 2a\rho(t) a^{\dag}-a^{\dag}a\rho(t)-\rho(t)a^{\dag}a$ represents the damping of the cavity field by cavity decay with the rate $\kappa$.

Using the master equation (\ref{e12}), we find equations of motion for the density matrix elements. 
For the initial conditions we choose the field to be in a coherent state with the amplitude~$\alpha$, and the atoms to be initially in a state determined by density matrix (\ref{e1}). By tracing over the cavity field, we obtain the reduced density matrix of the atoms, and find that the solutions for the density matrix elements are
\begin{eqnarray}
\rho_{11}(t) &=& \rho_{11}(0) ,\quad \rho_{44}(t) = \rho_{44}(0) ,\nonumber\\
\rho_{22}(t) &=& c_{+}(0) + c_{-}(0)\cos(\lambda{t}) - c_2(0)\sin(\lambda t) ,\nonumber\\
\rho_{33}(t) &=& 1 - \rho_{11}(t) - \rho_{22}(t) - \rho_{44}(t) ,\label{e14}\\
\rho_{23}(t) &=& c_{1}(0) + ic_{2}(0)\cos(\lambda{t}) + ic_{-}(0)\sin(\lambda{t}) ,\nonumber\\
\rho_{41}(t) &=& \rho_{41}(0)\exp\!\left\{-i\lambda t - \frac{2i\lambda|\alpha|^2}{\kappa + 2i\lambda}\!\left[1-{\rm e}^{-\left(\kappa+2i\lambda\right) t}\right]\!\right\} \nonumber ,
\end{eqnarray}
where $c_{\pm}(0) = (\rho_{22}(0)\pm\rho_{33}(0))/2$, and $\rho_{23}(0) = c_{1}(0)+ic_{2}(0)$. 
We see that the populations $\rho_{11}(t)$ and $\rho_{44}(t)$ are constants of motion that they retain their initial values for all times. Both $\rho_{22}(t)$ and $\rho_{33}(t)$ exhibit undamped Rabi oscillations. Similarly, the coherence $\rho_{23}(t)$ oscillates sinusoidally in time. Only the coherence $\rho_{14}(t)$ exhibits damped oscillations. Thus, the loss of the coherence $\rho_{23}(t)$ is reversible in time, it does not arise from the cavity damping mechanism. However, the loss of the coherence $\rho_{41}(t)$ is irreversible due to the damping of the cavity field. The coherence undergoes damped and phase sifted oscillations such that it decays to a non-zero stationary value 
\begin{eqnarray}
\lim_{t\rightarrow\infty}|\rho_{41}(t)| \rightarrow  |\rho_{41}(0)|\exp\!\left[-\frac{4\lambda^{2}|\alpha|^{2}}{\kappa^{2} +(4\lambda)^{2}}\right] .
\end{eqnarray}

To demonstrate that Eqs.~(\ref{e14}) can match the conditions for nullity of the quantum discord, we must compare the populations and the coherences. Since the populations $\rho_{11}(t)$ and $\rho_{44}(t)$ are constants of motion, the condition of $\rho_{11}(t) =\rho_{22}(t)$ and $\rho_{33}(t) =\rho_{44}(t)$ can be adjusted by a suitable choice of the initial atomic conditions. Thus, the problem simplifies to find if there are discrete times or periods of time at which the coherence $|\rho_{23}(t)| =|\rho_{14}(t)|$. Particularly interesting are the cases where $|\rho_{23}(t)| =|\rho_{14}(t)|\neq 0$ at $t\rightarrow \infty$, i.e., if the condition (\ref{e7}) can be satisfied in the steady state.

We now present some numerical calculations that illustrate the above remarks. We specialize to the time evolution of the quantum discord for different initial conditions to show that depending on the initial state, and also on the dissipation rate, the discord may never vanish, or may vanish at a finite number of discrete times, or even can vanish periodically in~time.
\begin{figure}[hpt]
\includegraphics[height=4cm,width=0.8\columnwidth]{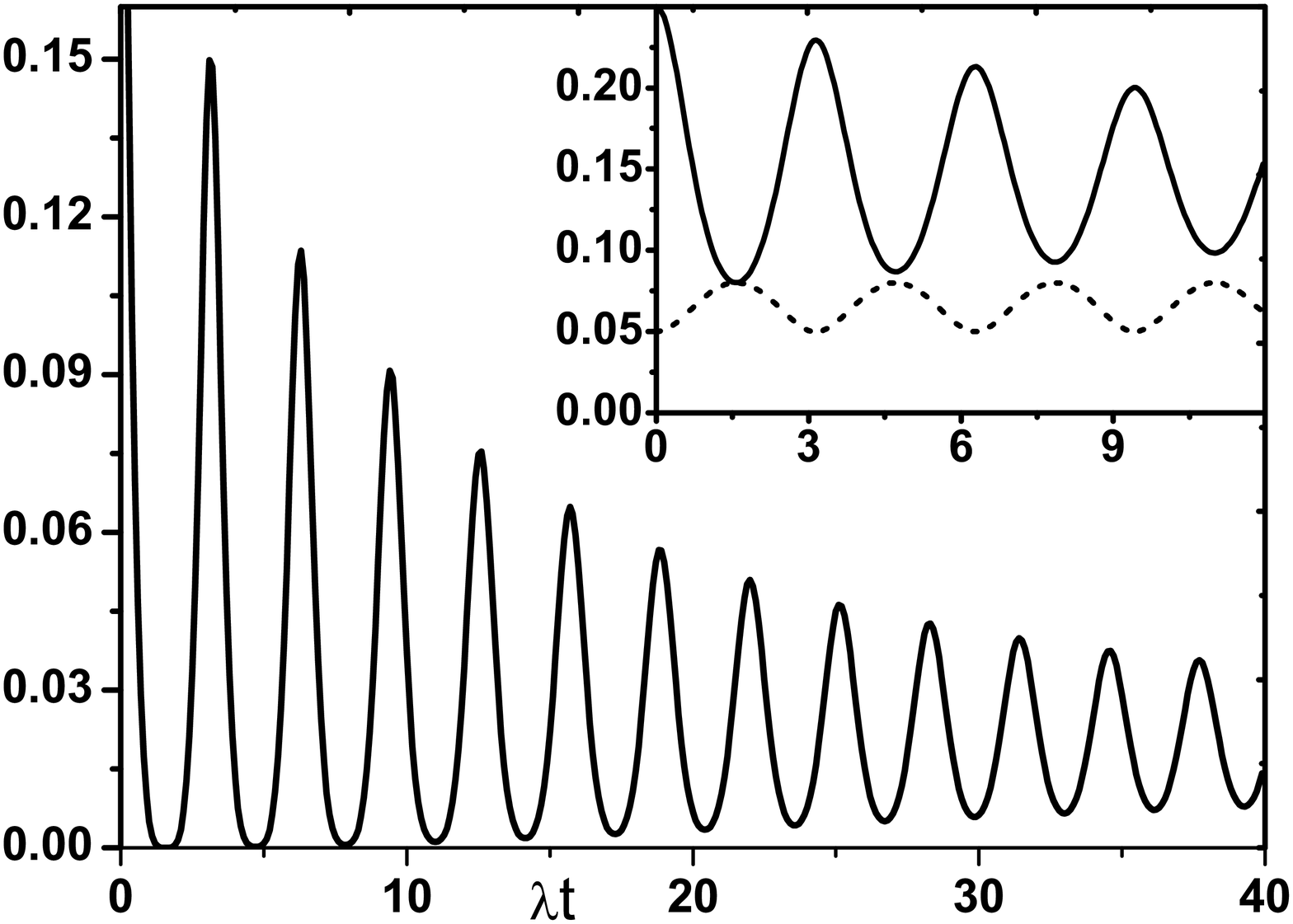}
\caption{Quantum discord (solid line) plotted as a function of dimensionless time $\lambda t$ for $|\alpha|^{2}= 0.5922$ and $\kappa = 0.05\lambda$. The system was initially in a state determined by $\rho_{11}(0)=\rho_{44}(0)= 1/4, \rho_{22}(0)=3/16, \rho_{33}(0)=5/16, |\rho_{23}(0)|= 0.05$, and $|\rho_{14}(0)|= 0.25$. The insert shows the time evolution of the quantum coherences $|\rho_{14}(t)|$ (solid line) and $|\rho_{23}(t)|$ (dashed line).}
\label{fig1}
\end{figure} 

Figure~\ref{fig1} shows the time evolution of the quantum discord for two qubits initially prepared in a non-zero discord state. One can see that the discord vanishes only at the first period of the oscillations when the coherences $\rho_{14}(t)$ and $\rho_{23}(t)$ merge. The coherences do not merge during any of the subsequent period of the oscillations. This simple example shows that a purely classical state or a pure classical nature of the system is a rather a short-lived affair. 

\begin{figure}[hpt]
\includegraphics[height=4cm,width=0.8\columnwidth]{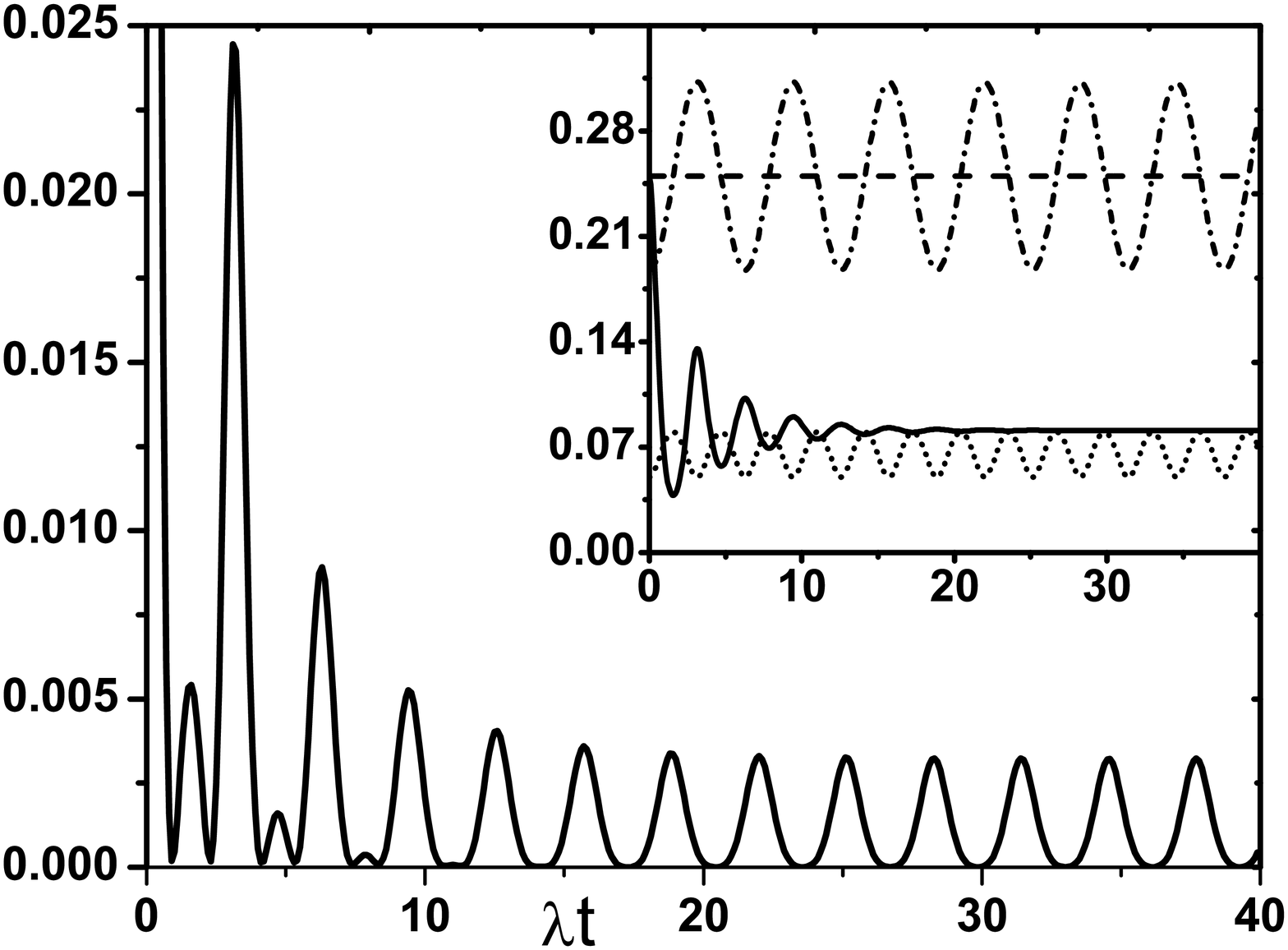}
\caption{Quantum discord (solid line) plotted as a function of dimensionless time $\lambda t$ for $|\alpha|^{2}= 1.1434$ and  $\kappa = 0.25\lambda$. The system was initially in a state determined by $\rho_{11}(0)=\rho_{44}(0)=1/4, \rho_{22}(0)=3/16 ,\rho_{33}(0)=5/16, |\rho_{23}(0)|=0.05$, and $|\rho_{14}(0)|=0.25$. The insert shows the time evolution of the populations $\rho_{22}(t)$ (dashed line) and $\rho_{33}(t)$ (dashed-dotted line), and also the quantum coherences $|\rho_{14}(t)|$ (solid line) and $|\rho_{23}(t)|$ (dotted line).}
\label{fig2}
\end{figure}

Figure~\ref{fig2} shows that a zero-discord state can be created not only by a suitable choice of the initial state, but also by a change of the cavity damping rate. We see that a large cavity damping rate results in a positive discord at short times with zeros occurring periodically at longer times. One can notice that at short times the coherences and the populations cross each other at some discrete times, so why the discord is positive at that times. The reason is that the coherences cross each other at times different than that the populations cross, so the condition (\ref{e7}) is not satisfied.

\begin{figure}[hpt]
\includegraphics[height=4cm,width=0.8\columnwidth]{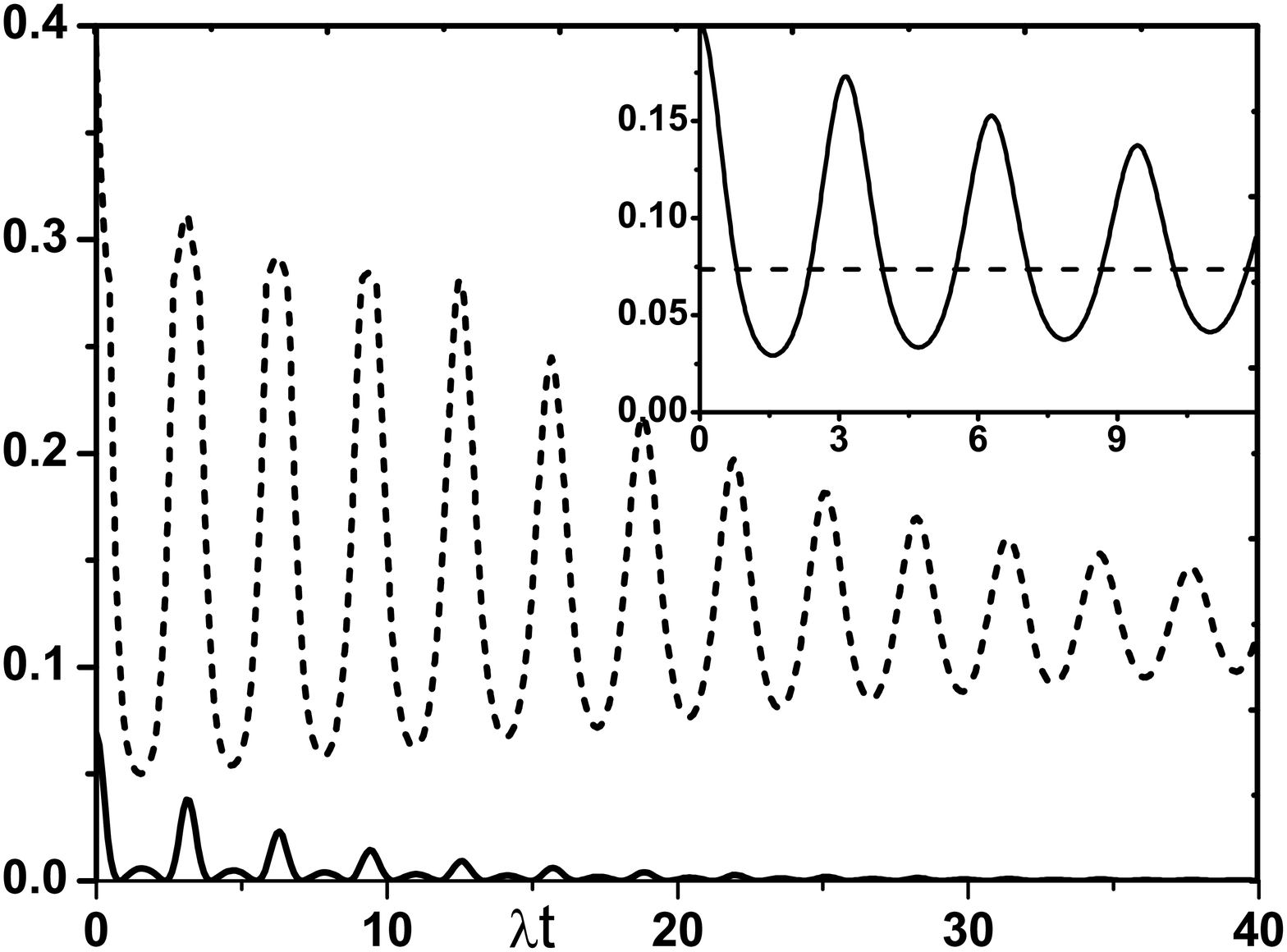}
\caption{Quantum discord plotted as a function of dimensionless time~$\lambda t$ for
$|\alpha|^{2}= 1.0$, $\kappa = 0.05\lambda$ and two different initial Bell-diagonal states; $\rho_{jj}(0)=1/4, |\rho_{23}(0)|=0.0736, |\rho_{14}(0)|=0.2$ (solid line), and $\rho_{11}(0)=\rho_{44}(0)=|\rho_{14}(0)|=0.4, \rho_{22}(0)=\rho_{33}(0)=0.1, |\rho_{23}(0)|=0.05$ (dashed line). The insert shows the time evolution of the quantum coherences $|\rho_{14}(t)|$ (solid line) and $|\rho_{23}(t)|$ (dashed line) for the case of the initially unentangled qubits.}
\label{fig3}
\end{figure}

It is interesting that the possibility for the discord to vanish depends on whether the initial state was entangled or not. This is illustrated in Fig.~\ref{fig3}, where the plot the time evolution of the discord for two different initial Bell-diagonal states~\cite{ds08}, mixed states whose eigenstates are four maximally entangled states, the Bell states. For the initial unentangled state, the discord vanishes periodically in time due to a modulation of the Rabi oscillations. The modulation results from the crossing of the coherences periodically at discrete times. The coherence $|\rho_{23}|$ is constant in time and, as the time progresses, the coherence $|\rho_{14}|$ evolves towards its non-zero stationary value of $|\rho_{14}| =|\rho_{23}|$. In contrast, the discord decreases in time and vanishes at the steady state. Note that the collapse of the Rabi oscillations of the coherence $|\rho_{14}|$ is far from complete when the discord disappears. The behaviour of the discord is different for the initial entangled state. In this case, the discord differs from zero for all times that makes us to conclude that the initial entangled state wipes out zeros of the discord. Hence, we may state that quantum correlations of an initially entangled Bell-diagonal state cannot be completely destroyed.

Now, we briefly comment about the evolution of the quantum discord from an initial zero-discord state. In this case, the phase damping of the coherence $\rho_{14}(t)$ breaks the condition of $|\rho_{14}(t)| = |\rho_{23}(t)|$. Thus the phase  damping, it turns out, is sufficient to wipe out any zero-discord state. In other words, the dissipation can build up quantum correlations, leaving purely classical states to occur only during the initial time. 

Finally, we discuss a possible experimental arrangement where the zero-discord states could be measured. A good candidate is the scheme used by Osnaghi {\it et al.}~\cite{ob01} to observe entanglement between two Rydberg atoms crossing a nonresonant microwave cavity. The scheme involves the Tavis-Cummings model determined by the Hamiltonian of the same form as Eq.~(\ref{e11}). Therefore, we anticipate no difficulty to modify the scheme for the measurement of the discord. 

{\it In summary}, we have derived conditions for nullity of quantum discord of an arbitrary two-qubit  system whose dynamical behaviour is given by an $X$-state form density matrix. The results derived show that the quantum discord can vanish even in the presence of quantum coherences between the qubits. The condition is related physically to the preparation of the spin correlated and anti-correlated states in equal superposition states. We have shown that the condition can be realized in a practical system of the coherent-state Tavis-Cummings model in which two distant qubits interact dispersively with the cavity field.

This work was supported by the National Natural Science Foundation of China (Grant Nos. 60878004
and 11074087), the Ministry of Education under project SRFDP (Grant No. 200805110002), and the Natural Science Foundation of Hubei Province (Grant No. 2010CDA075).

\end{document}